\documentclass[12pt,preprint]{aastex}
\begin{document}
\title{On the Growth of Perturbations as a Test of Dark Energy and Gravity}
\author{Edmund Bertschinger}
\affil{Department of Physics and Kavli Institute for Astrophysics
and Space Research, MIT Room 37-602A, 77 Massachusetts Ave.,
  Cambridge, MA 02139; \email{edbert@mit.edu}}
\begin{abstract}
The strongest evidence for dark energy comes presently from
geometric techniques such as the supernova distance-redshift
relation.  By combining the measured expansion history with the
Friedmann equation one determines the energy density and its time
evolution, hence the equation of state of dark energy. Because these
methods rely on the Friedmann equation which has not been
independently tested it is desirable to find alternative methods
that work for both general relativity and other theories of gravity.

Assuming that sufficiently large patches of a perturbed
Robertson-Walker spacetime evolve like separate Robertson-Walker
universes, that shear stress is unimportant on large scales and that
energy and momentum are locally conserved, we derive several
relations between long-wavelength metric and matter perturbations.
These relations include generalizations of the initial-value
constraints of general relativity. For a class of theories including
general relativity we reduce the long-wavelength metric, density,
and velocity potential perturbations to quadratures including
curvature perturbations, entropy perturbations, and the effects of
nonzero background curvature.  When combined with the expansion
history measured geometrically, the long-wavelength solution provide
a test that may distinguish modified gravity from other explanations
of dark energy.
\end{abstract}

\section{Introduction}

Current evidence for dark energy is based on two key assumptions:
\begin{enumerate}
  \item The Cosmological Principle holds, i.e. on large scales
  the matter distribution and its expansion are homogeneous and
  isotropic and the spacetime geometry is Robertson-Walker;
  and
  \item The cosmic expansion scale factor $a(t)$ obeys the
  Friedmann equation, which may be written
  \begin{equation}\label{friedmann}
    {\cal H}^2\equiv\left(\frac{1}{a}\frac{da}{d\tau}\right)^2=a^2H^2=
    \frac{8\pi Ga^2}{3}\rho(a)-K\  ,
  \end{equation}
\end{enumerate}
where $K$ is the spatial curvature with units of inverse length
squared (we set $c=1$) and $\tau$ is conformal time, related to
cosmic proper time $t$ by $dt=a(\tau)d\tau$.  Usually the
Friedmann equation is applied indirectly through an expression for
the angular-diameter or luminosity distance; these formulae depend
crucially on the expression for $\tau(a)$ obtained by integrating
equation (\ref{friedmann}).

The Cosmological Principle is more general than general relativity
(GR) and it is amenable to direct observational test through
measurements of distant objects such as galaxies, Type Ia
supernovae, and the microwave background radiation. The Friedmann
equation, however, is equivalent to one of the Einstein field
equations of GR applied to the Robertson-Walker metric and, so far
at least, it has not been independently tested. Instead, the
Friedmann equation is used by astronomers in effect to deduce
$\rho(a)$ including dark energy through measurements of the Hubble
expansion rate $H(a)$.

If the evidence for dark energy is secure, there are four possible
explanations:
\begin{enumerate}
  \item The dark energy is a cosmological constant or,
  equivalently, the energy density and negative pressure of the vacuum
  (Gliner 1966; Zel'dovich 1967).
  \item The dark energy is some other source of stress-energy, for
  example a scalar field with large negative pressure (Ratra \&
  Peebles 1988; Steinhardt et al.\ 1999).
  \item General relativity needs to be modified and we cannot use
  it to deduce the existence of either a cosmological constant or
  an exotic form of energy (Dvali et al.\ 2000; Lue et al.\ 2004).
  \item General relativity is correct but our understanding of it
  is not; e.g.\ long-wavelength density perturbations modify the
  Friedmann equation (Kolb et al.\ 2005a,b).
\end{enumerate}

It would be interesting to find observational tests that can
distinguish among these possibilities.  The first case can be tested
by measuring the equation of state parameter $w\equiv p/\rho$ on
large scales; a cosmological constant has $w=-1$.  This test can be
made using methods such as the supernova distance-redshift relation
(Riess et al.\ 1998; Perlmutter et al.\ 1999) and baryon acoustic
oscillations (Eisenstein et al.\ 2005) whose interpretation relies
on the Friedmann equation. Such methods are called geometric (they
measure the large-scale geometry of spacetime) or kinematic (they
rely on energy conservation and on the first time derivative of the
expansion scale factor).

It is more difficult to test the third possibility, namely that dark
energy represents a modification of GR rather than (or in addition
to) a new form of mass-energy. Redshift-distance tests invoke the
Friedmann equation of GR or its equivalent in other theories in
order to determine the abundance and equation of state of dark
energy. Such tests do not work without a formula relating $a(\tau)$
to the cosmic energy density and pressure.  While specific
alternative models can be tested, it would be desirable to have
general cosmological tests independent of the Friedmann equation.

The evolution of density perturbations has been proposed as an
independent test of dark energy and general relativity (Linder 2005;
Ishak et al.\ 2005). For example, weak gravitational lensing
measurements, the evolution of galaxy clustering on large scales,
and the abundance of rich galaxy clusters all have some sensitivity
to the gravitational effects of dark energy at redshifts $z<1$.
These methods may be called dynamic because the evolution equations
for the perturbations are at least second order in time.

This paper examines what the linear growth of metric, density, and
velocity perturbations can tell us about gravity. By generalizing
previous work on evolution of separate universes (Wands et al.\
2000; Gordon 2005 and references therein), we will show that in GR,
on length scales larger than the Jeans length (more generally, on
scales large enough so that spatial gradients may be neglected in
the equations of motion), the evolution of the metric and density
perturbations of a background Robertson-Walker universe can be
determined from the Friedmann equation and local energy-momentum
conservation. Generalizing this to arbitrary theories of gravity,
under certain conditions explained in this paper, the evolution of
Robertson-Walker spacetimes (assuming they are solutions of the
gravitational field equations) combined with local energy-momentum
conservation is sufficient to determine the evolution of
long-wavelength perturbations of the metric and matter variables.

Because GR is so fully integrated into most treatments of
cosmological perturbation theory, and we wish to test gravitation
more generally, it is worth recalling its basic elements:
\begin{enumerate}
  \item Spacetime is describable as a classical 4-dimensional
    manifold with a metric locally equivalent to Minkowski.  This
    is generalizable to higher dimensions where matter fields reside
    on a three-dimensional spatial brane.
  \item Special relativity holds locally.  In particular, energy-momentum
    is locally conserved.  This is more general than GR.
  \item The weak equivalence principle holds, i.e. freely-falling
    bodies follow spacetime geodesics.  This is more general then
    GR.
  \item The metric is the solution to the Einstein field equations
    subject to appropriate initial and boundary conditions.  This is
    uniquely true in GR.
\end{enumerate}
The first three ingredients are assumed by most viable theories of
gravitation applied on cosmological scales.  They are assumed to be
correct throughout this paper. The Einstein field equations are not
assumed to hold except where explicitly stated below. We assume
throughout that the universe is approximately (or, for some
calculations, exactly) Robertson-Walker.

\section{Robertson-Walker Spacetimes and their Perturbations}

The general Robertson-Walker spacetime is specified by giving a
spatial curvature constant $K$ (with units of inverse length
squared) and a dimensionless scale factor $a(\tau)$ (normalized so
that $a=1$ today). Let us assume that the evolution of $a(\tau)$
depends on $K$ and on the properties of the matter and energy
filling the universe. Applying the Cosmological Principle, the
matter must behave as a perfect fluid at rest in the comoving frame.
The pressure of a perfect fluid may be written $p(\rho,S)$ where
$\rho$ is the proper energy density and $S$ is the comoving entropy
density in the fluid rest frame. The line element may thus be
written
\begin{equation}\label{rw}
  ds^2=a^2(\tau,K,S)\left[-d\tau^2+d\chi^2+r^2(\chi,K)d\Omega^2\right]
\end{equation}
where $r(\chi,K)=K^{-1/2}\sin(K^{1/2}\,\chi)$ for $K>0$ (and is
analytically continued for $K\le0$) and $d\Omega^2\equiv
d\theta^2+\sin^2\theta\,d\phi^2$. Initially we make no assumption
about the dynamics except that the evolution of the scale factor
depends only on the geometry ($K$) and composition ($S$) of the
(3+1)-dimensional universe and not, for example, on parameters
describing extra dimensions.\footnote{It would be straightforward to
add such parameters to the argument list of $a(\tau,K,S)$ and then
perturb them.} Only later will the Friedmann equation be assumed to
determine the exact form of $a(\tau,K,S)$.

\subsection{Curvature and Coordinate Perturbations}
\label{sec:cpert}

Metric perturbations are obtained by comparing two slightly
different spacetimes.  We consider two homogeneous and isotropic
Robertson-Walker spacetimes differing only by their (spatially
homogeneous) spatial curvature.  The first spacetime has spatial
curvature $K$; the second one has spatial curvature $K(1+\delta_K)$
where $\delta_K$ is a small constant.  We write the metric of the
second spacetime as a perturbation of the first, as follows.  First,
the angular radius may be Taylor-expanded to first order in
$\delta_K$ to give
\begin{equation}\label{modr2}
  r^2(\chi,K+K\delta_K)=
    r^2(\chi,K)(1-\delta_K)+\chi r(\chi,4K)\delta_K\ ,
\end{equation}
while $a(\tau,K+K\delta_K,S)=a(\tau,K,S)+\delta_K(\partial a/
\partial\ln K)$. To simplify the appearance of the line element we
change variables $\tau\to\tau+\alpha(\tau)$ and
$\chi\to\chi(1-\kappa)$ where $\alpha$ and $\kappa$ are assumed to
be first order in $\delta_K$. In these new coordinates, to first
order in $\delta_K$ the second spacetime has line element
\begin{eqnarray}\label{rw2}
  ds^2&=&a^2(\tau,K,S)\left(1+2{\cal H}\alpha
    +2\delta_K\frac{\partial\ln a}{\partial\ln K}\right)\\
  &&\times\left\{-(1+2\dot\alpha)d\tau^2
    +(1-2\kappa)d\chi^2+\left[(1-\delta_K)r^2(\chi,K)
    +(\delta_K-2\kappa)\chi r(\chi,4K)\right]d\Omega^2\right\}\ ,
    \nonumber
\end{eqnarray}
where ${\cal H}(\tau,K,S)\equiv\partial\ln a/\partial\tau$. This
line element describes a perfectly homogeneous and isotropic
Robertson-Walker spacetime with spatial curvature $K(1+\delta_K)$.
However, for appropriate choices of $\alpha$ and $\kappa$ (i.e.,
the appropriate coordinate transformation), it takes precisely the
same form as a perturbed Robertson-Walker spacetime with
background spatial curvature $K$ and perturbation $\Psi$,
\begin{equation}\label{rw3}
  ds^2=a^2(\tau,K,S)\left\{-(1+2\Psi)d\tau^2+(1-2\Psi)\left[d\chi^2
    +r^2(\chi,K)d\Omega^2\right]\right\}\ ,
\end{equation}
provided that the following three conditions hold:
\begin{equation}\label{rwmatch1}
  \kappa=\frac{1}{2}\delta_K\ ,\ \ 2\Psi=\dot\alpha+\kappa\ ,\ \
  \Psi=\kappa\left(1-2\frac{\partial\ln a}{\partial\ln K}\right)
    -{\cal H}\alpha\ .
\end{equation}
In other words, a perturbed Robertson-Walker spacetime whose metric
perturbation $\Psi(\tau)$ is spatially homogeneous is identical to
an unperturbed Robertson-Walker spacetime represented by a perturbed
coordinate system.\footnote{In \S \ref{sec:alt} we will generalize
equation (\ref{rw3}) to the case of two distinct potentials for the
time- and space- parts of the metric.}

This equivalence is significant because it suggests that
long-wavelength curvature perturbations (for which $\Psi$ is
effectively independent of spatial position) should evolve like
patches of a Robertson-Walker spacetime (Wands et al.\ 2000; Gordon
2005 and references therein), whose dynamics is more general than
general relativity. So far we have assumed that the Cosmological
Principle holds but we have not assumed the validity of the Einstein
field equations of general relativity.  However, if we know how a
perfect Robertson-Walker spacetime evolves, the evolution of
long-wavelength curvature perturbations follows.

To make further progress let us assume the form of $a(\tau,K,S)$,
which requires specifying a theory of gravity.  The simplest choice
is to assume the validity of the Friedmann equation, which gives
\begin{equation}\label{friedsol}
  \tau(a,K,S)=\int_0^a\left[\frac{8\pi}{3}G\tilde a^4\rho(\tilde a,S)
    -K\tilde a^2\right]^{-1/2}\,d\tilde a\ .
\end{equation}
Using this, one finds
\begin{equation}\label{aderivK}
  \left(\frac{\partial\ln a}{\partial\ln K}\right)_{\tau,S}
    =-\frac{(\partial\tau/\partial\ln K)_{a,S}}{(\partial\tau/
    \partial\ln a)_{K,S}}=-\frac{1}{2}K{\cal H}
    \int^\tau\frac{d\tau'}{{\cal H}^2(\tau',K,S)}\ .
\end{equation}
We also require conservation of energy, which may be written
\begin{equation}\label{econs}
  \frac{\partial}{\partial\ln a}\rho(a,S)=-3[\rho+p(\rho,S)]\ .
\end{equation}
When combined with the Friedmann equation, this implies
\begin{equation}\label{Hdot}
  \gamma\equiv4\pi Ga^2(\rho+p)={\cal H}^2+K-\dot{\cal H}
  =\frac{3}{2}(1+w)({\cal H}^2+K)\ ,
\end{equation}
where $\dot{\cal H}=\partial{\cal H}(\tau,K,S)/\partial\tau$.
Combining equations (\ref{rwmatch1}), (\ref{aderivK}), and
(\ref{Hdot}) yields a relation between the long-wavelength potential
$\Psi$ and the curvature perturbation\footnote{For spatially
uniform, isentropic perturbations with vanishing shear stress,
$\kappa$ reduces to the $\zeta$ variable of Bardeen et al. (1983).
In other cases the two variables differ, with $\kappa$ being simpler
in both its dynamics and interpretation.} $\kappa$:
\begin{equation}\label{psikappa}
  \kappa=\frac{{\cal H}^2}{\gamma a^2}\frac{\partial}{\partial\tau}
    \left(\frac{a^2\Psi}{{\cal H}}\right)\ .
\end{equation}
For long wavelengths $\kappa$ may depend on the wavenumber $k$ but
cannot depend on $\tau$, so $\partial\kappa/\partial\tau=0$, which
implying
\begin{equation}\label{psiddotfrw}
  \frac{\gamma}{{\cal H}}\frac{\partial}{\partial\tau}\left[\frac{{\cal H}^2}
    {\gamma a^2}\frac{\partial}{\partial\tau}\left(\frac{a^2\Psi}{{\cal H}}
    \right)\right]=\ddot\Psi+3(1+c_w^2){\cal H}\dot\Psi+3(c_w^2-w){\cal H}^2
    \Psi-(2+3w+3c_w^2)K\Psi=0\ ,
\end{equation}
where
\begin{equation}\label{eos}
  w\equiv\frac{p(\rho,S)}{\rho}\ ,\ \ c_w^2\equiv\left(\frac{\partial p}
    {\partial\rho}\right)_{\!S}\ .
\end{equation}

By comparing two different Friedmann-Robertson-Walker models we have
arrived at a second-order differential equation for the metrc
perturbation $\Psi$. The only dynamical equations assumed have been
the Friedmann equation and energy conservation in a homogeneous and
isotropic universe.  We have not made use of the perturbed Einstein
or fluid equations.  Nevertheless, as we show next, equation
(\ref{psiddotfrw}) is identical with the dynamical evolution
equation for long-wavelength curvature perturbations obtained using
the perturbed fluid and Einstein equations of general relativity.

\subsection{Linear Cosmological Perturbations in General
Relativity}

Cosmological perturbation theory has been well studied (e.g.\
Lifshitz 1946; Bardeen 1980; Kodama \& Sasaki 1984; Hwang \& Noh
2002). Nonetheless, the curvature and entropy variables relevant for
long-wavelength density perturbations differ from those given
previously in the literature, so a brief summary is presented here.

In the conformal Newtonian gauge (Mukhanov et al.\ 1992; Ma \&
Bertschinger 1995), the metric of a perturbed Robertson-Walker
spacetime with scalar perturbations may be written as a
generalization of equation (\ref{rw3}),
\begin{equation}\label{pertrw}
  ds^2=a^2(\tau)\left\{-(1+2\Phi)d\tau^2+(1-2\Psi)
    [d\chi^2+r^2(\chi,K)d\Omega^2]\right\}\ .
\end{equation}
Here, $\Phi(x^i,\tau)$ and $\Psi(x^i,\tau)$ are small-amplitude
gravitational potentials.  The dependence of the scale factor on $K$
and $S$ is suppressed because we are considering now a single
universe with unique values of these parameters and are introducing
perturbations only through the potentials.

For scalar perturbations the stress-energy tensor components may
be written in terms of spatial scalar fields $\delta\rho$, $u$,
and $\pi$, as follows,
\begin{mathletters}\label{emtensor}
\begin{eqnarray}
  T^0_{\ \,0}&=&-(\bar\rho+\delta\rho)\ ,\ \label{t00}\\
  T^0_{\ \,i}&=&-(\bar\rho+\bar p)\nabla_i u\ ,\label{t0i}\\
  T^i_{\ \,j}&=&\delta^i_{\ \,j}(\bar p+\delta p)+\frac{3}{2}
    (\bar\rho+\bar p)\left(\nabla^i\nabla_j-\frac{1}{3}\delta^i_{\ \,j}
     \Delta\right)\pi\ ,\label{tij}
\end{eqnarray}
\end{mathletters}
where $\nabla_i$ and $\nabla^i$ are the three-dimensional covariant
derivative for the spatial line element $d\chi^2+r^2d\Omega^2$ while
$\Delta=\nabla^i\nabla_i$. Unless stated otherwise, all variables
refer to the total stress-energy summed over all components. The
unperturbed energy density and pressure are $\bar\rho(\tau)$ and
$\bar p(\tau)$, respectively, while $\delta\rho$ and $\delta p$ are
the corresponding perturbations measured in the coordinate frame.

In the scalar mode, all perturbations to the metric and the
stress-energy tensor arise from spatial scalar fields and their
spatial gradients.  For example, the energy flux is a potential
field with velocity potential $u(x^i,\tau)$. Similarly, the shear
stress follows from a shear stress (viscosity) potential
$\pi(x^i,\tau)$ (defined as in Bashinsky \& Seljak 2004). For an
ideal gas, $\pi=0$. Equations (\ref{emtensor}) are completely
general for the scalar mode. Stress-energy perturbations that arise
from divergenceless (transverse) vectors contribute only to the
vector mode, while divergenceless, trace-free tensors contribute
only to the tensor mode.  Vector and tensor modes are ignored in
this paper.

Combining the perturbed Einstein field equations (Bertschinger 1996)
with equations (\ref{Hdot}) and (\ref{eos}) gives the following
second-order partial differential equation for the linear evolution
of the potential $\Psi$:
\begin{eqnarray}\label{psieom}
  \frac{\gamma}{{\cal H}}\frac{\partial}{\partial\tau}\left[
    \frac{{\cal H}^2}{\gamma a^2}\frac{\partial}{\partial\tau}
    \left(\frac{a^2}{{\cal H}}\Psi\right)\right]-c_w^2\Delta
    \Psi
  =\gamma\left[\frac{\delta p-c_w^2\delta\rho}{\bar\rho
    +\bar p}+\frac{3}{{\cal H}}\frac{\partial}{\partial
    \tau}\left({\cal H}^2\pi\right)+\Delta\pi\right]\ .
\end{eqnarray}
Equation (\ref{psieom}) is exact in linear perturbation theory of
general relativity and is fully general for scalar perturbations. It
was given in another form by Hwang \& Noh (2002). Aside from the
right-hand side and the sound speed term $c_w^2\Delta\Psi$, it
agrees exactly with equation (\ref{psiddotfrw}), whose derivation
did not assume the validity of the perturbed Einstein field
equations. The right-hand side of equation (\ref{psieom}) includes
entropy perturbations proportional to $\delta p-c_w^2\delta\rho$ and
shear stress perturbations, both of which were absent in the
Robertson-Walker models of the previous subsection.

In a perfectly homogeneous and isotropic universe the shear stress
must vanish, and a correct treatment of shear stress requires going
beyond the Friedmann and energy conservation equations.  In the
universe at low redshift, the shear stress due to photons,
neutrinos, and gravitationally bound structures is orders of
magnitude smaller than the mass density perturbations. Unless the
dark energy is a peculiar substance with large shear stress, we may
neglect $\pi$ in the equation of motion for $\Psi$.

The sound wave term represents the effect of pressure forces in
resisting gravitational instability.  For a perturbation of comoving
wavenumber $k$, $-c_w^2\Delta\Psi=k^2c_w^2\Psi$; for comparison the
time derivative terms in equation (\ref{psieom}) are $\sim {\cal
H}^2\Psi$. Thus, for wavelengths much longer than the comoving Jeans
length $\lambda_J$ defined by $\pi/\lambda_J={\cal H}/c_w$, the
sound wave term may be neglected.  In the standard cosmology, the
Jeans length at $z<100$ is less than about 20 Mpc.

The entropy source term in equation (\ref{psieom}) can also be
obtained by comparing separate Robertson-Walker spacetimes with
slightly different entropies, allowing us to reduce the
long-wavelength evolution entirely to quadratures, as we show next.

\subsection{Entropy Perturbations}
\label{sec:epert}

Consider two Robertson-Walker spacetimes with identical spatial
curvature $K$ but with entropies $S$ and $S+\delta S$ respectively,
where $\delta S$ is a small constant. At a given expansion factor
$a$ ($a^3$ plays the role of volume), the density $\rho(a,S)$ and
pressure $p(\rho,S)$ will differ slightly in the two spacetimes.
Writing the pressure as $p(\rho(a,S),S)$, from equations
(\ref{econs}) and (\ref{eos}) we obtain
\begin{equation}\label{pderivs}
  \left(\frac{\partial p}{\partial\ln a}\right)_S=-3c_w^2(\rho+p)
    \ ,\ \
  \left(\frac{\partial p}{\partial S}\right)_{\!a}
    =\left(\frac{\partial p}{\partial S}\right)_{\!\rho}
    +c_w^2\left(\frac{\partial\rho}{\partial S}\right)_{\!a}\ .
\end{equation}
Using equations (\ref{econs}) and (\ref{pderivs}), we find
\begin{equation}\label{psderiv}
  \left(\frac{\partial p}{\partial S}\right)_{\!\rho}
  =-\left(\frac{\rho+p}{3}\right)\frac{\partial}{\partial\ln a}
    \left[\frac{1}{\rho+p}\left(\frac{\partial\rho}{\partial S}
    \right)_{\!a}\,\right]\ .
\end{equation}
Using this result, we define a fractional entropy perturbation
variable
\begin{equation}\label{sigdef}
  \sigma(a,S)\equiv\frac{\delta p-c_w^2\delta\rho}{\rho+p}
  =\frac{\delta S}{\rho+p}\left(\frac{\partial p}{\partial S}
    \right)_{\!\rho}=-\frac{\delta S}{3}\frac{\partial}{\partial
    \ln a}\left[\frac{1}{\rho+p}\left(\frac{\partial\rho}{\partial S}
    \right)_{\!a}\,\right]\ .
\end{equation}
This is a formal result because for a multicomponent imperfect fluid
one would not evaluate $\rho(a,S)$ but would instead characterize
density and pressure perturbations for the individual components, as
we shown below in \S \ref{sec:multi}.  For now, we assume that
$\sigma$ can be determined and we use it to derive a quadrature for
the isocurvature modes.

The scale factor in the second Robertson-Walker spacetime is
$a(\tau,K,S+\delta S)=a(\tau,K,S)+(\partial a/\partial S)\delta S$.
Changing time variable $\tau\to\tau+\alpha(\tau)$ (with no change in
spatial coordinates), the line element for the second spacetime
becomes
\begin{equation}\label{rw4}
  ds^2=a^2(\tau,K,S)\left(1+2{\cal H}\alpha+2\delta S\frac
    {\partial\ln a}{\partial S}\right)\left[-(1+2\dot\alpha)
    d\tau^2+d\chi^2+r^2(\chi,K)d\Omega^2\right]\ .
\end{equation}
This line element describes a perfectly homogeneous and isotropic
Robertson-Walker spacetime with entropy $S+\delta S$. However, for
appropriate choice of $\alpha$ (i.e., the appropriate coordinate
transformation), it takes precisely the same form as a perturbed
Robertson-Walker spacetime with background entropy $S$, given by
equation (\ref{rw3}), provided that the following two conditions
hold:
\begin{equation}\label{rwmatch2}
  2\Psi=\dot\alpha\ ,\ \
  \Psi=-\delta S\frac{\partial\ln a}{\partial S}-{\cal H}\alpha\ .
\end{equation}

So far we have assumed nothing about gravity.  To proceed further we
assume that equation (\ref{friedsol}) is valid, giving
\begin{equation}\label{aderivS}
  \left(\frac{\partial\ln a}{\partial S}\right)_{\tau,K}
    =-\frac{(\partial\tau/\partial S)_{a,K}}{(\partial\tau/
    \partial\ln a)_{K,S}}=\frac{4\pi G{\cal H}}{3}\int^\tau
    \frac{a^2(\tau',K,S)}{{\cal H}^2(\tau',K,S)}\left(
    \frac{\partial\rho}{\partial S}\right)_{\!a'}\,d\tau'\ .
\end{equation}
Combining equations (\ref{Hdot}), (\ref{sigdef}), (\ref{rwmatch2})
and (\ref{aderivS}) gives
\begin{eqnarray}\label{psidotfrw}
  \frac{1}{a^2}\frac{\partial}{\partial\tau}\left(\frac{a^2\Psi}
    {{\cal H}}\right)&=&-\delta S\frac{\partial}{\partial\tau}
    \left[\frac{1}{{\cal H}}\left(\frac{\partial\ln a}{\partial S}
    \right)_{\!\tau,K}\,\right]
  =-\frac{4\pi Ga^2\delta S}{3{\cal H}^2}\left(\frac
    {\partial\rho}{\partial S}\right)_{\!a}\nonumber\\
  &=&\frac{\gamma}{{\cal H}^2}
    \int^\tau\sigma(a(\tau',K,S),S){\cal H}(\tau',K,S)\,d\tau'\ .
\end{eqnarray}
This result agrees exactly with equation (\ref{psieom}) for
long-wavelength entropy perturbations in GR with vanishing shear
stress potential $\pi$.  Therefore, in GR the dynamical evolution of
long-wavelength entropy perturbations in a Robertson-Walker
spacetime follows directly from the Friedmann and energy
conservation equations without requiring the perturbed Einstein
field equations. Equation (\ref{psidotfrw}) implies equation
(\ref{psiddotfrw}) when the entropy perturbation vanishes.

\subsection{General Solution for Long-Wavelength Perturbations in GR}
\label{sec:quadGR}

Equations (\ref{psiddotfrw}) and (\ref{psidotfrw}) have a simple
exact solution. The homogeneous solution with $\sigma=0$ (curvature
perturbations) is (suppressing the dependencies on $K$ and $S$,
which are held fixed)
\begin{equation}\label{curv}
  \Psi(\tau)=\kappa\Psi_+(\tau)+C\Psi_-(\tau)\ ,\ \
  \Psi_+(\tau)=\frac{{\cal H}}{a^2}\int^\tau\frac{\gamma(\tau')
    a^2(\tau')}{{\cal H}^2(\tau')}\,d\tau'\ ,\ \
    \Psi_-(\tau)=\frac{{\cal H}}{a^2}\ .
\end{equation}
Here $\kappa$ is the curvature perturbation of equation
(\ref{psikappa}) and it gives the amplitude of the ``growing mode''
of density perturbations; because the lower limit of integration for
$\Psi_+(\tau)$ is unspecified one may add any constant multiple $C$
of the ``decaying mode'' solution $\Psi_-(\tau)={\cal H}a^{-2}$. The
decaying mode is a gauge mode which can be eliminated using the
coordinate transformation $\tau\to\tau+Ca^{-2}$. The particular
solution with $\Psi=\dot\Psi=0$ at $\tau=0$ (isocurvature
perturbations) and $\sigma\ne0$ (again suppressing the dependencies
on $K$ and $S$) is
\begin{equation}\label{isocurv}
  \Psi(\tau)=\int_0^\tau\left[\Psi_+(\tau)\Psi_-(\tau')-\Psi_+(\tau')
    \Psi_-(\tau)\right]\sigma(\tau')a^2(\tau')\,d\tau'\ .
\end{equation}

The general solution of equation (\ref{psieom}) with $\pi=\Delta
\Psi=0$ is given by adding equations (\ref{curv}) and
(\ref{isocurv}).  We see that it is equivalent to integration of
\begin{equation}\label{rwmatch3}
  2\Psi=\dot\alpha+\kappa\ ,\ \
  \Psi=-{\cal H}\alpha+\kappa\left(1-2\frac{\partial\ln a}
    {\partial\ln K}\right)-\delta S\frac{\partial\ln a}{\partial S}
\end{equation}
when the Friedmann equation governs the background expansion. By
comparing the evolution of separate Robertson-Walker universes we
have solved equations (\ref{rwmatch3}) by the quadratures of
equations (\ref{curv}) and (\ref{isocurv}).  The solution requires
only that the background evolution obeys the Friedmann and energy
conservation equations, and that spatial gradient terms are
negligible. Later we will drop the assumption of the Friedmann
equation to obtain quadratures for any theory of gravity.

Although we have set $k^2=-\Delta=0$, the solution is valid for all
wavelengths much greater than the Jeans length.  One simply allows
$\kappa$ and $\delta S$ to depend on wavevector $\vec k$ as
determined by initial conditions.  At short wavelengths (shorter
than the Jeans length, for example) $\kappa$ and $\delta S$, as
obtained using equations (\ref{psikappa}) and (\ref{sigdef}), depend
on time.

\section{Density and Velocity Perturbations and Einstein
Constraints}

In addition to the metric perturbations, the density and velocity
perturbations of the matter can also be reduced to quadratures in
the long-wavelength limit assuming local energy-momentum
conservation.  The density and velocity perturbations then obey
initial-value constraint equations which relate them to the metric
perturbations.

Under the coordinate transformations given in the preceding
sections, the value of $T^0_{\ \,0}$ does not change to first order
in the perturbations.  However, comparing the density field of two
different Robertson-Walker spacetimes at the same coordinate values
$(\tau,x^i)$ gives a density perturbation:\footnote{Note the unusual
definition of $\delta$; for a gas of particles it gives the relative
number density perturbation rather than the relative energy density
perturbation.  The equations simplify with this choice.}
\begin{eqnarray}\label{denpert}
  \delta(\tau)\equiv\frac{\delta\rho}{\bar\rho+\bar p}&=&-3\left[{\cal H}
    \alpha+2\kappa\left(\frac{\partial\ln a}{\partial\ln K}\right)_{\tau,S}
    +\delta S\left(\frac{\partial\ln a}{\partial S}\right)_{\tau,K}
    \right]+\frac{\delta S}{\bar\rho+\bar p}\left(
    \frac{\partial\rho}{\partial S}\right)_{\!a}\nonumber\\
  &=&3(\Psi-\kappa)+\frac{\delta S}{\bar\rho+\bar p}\left(
    \frac{\partial\rho}{\partial S}\right)_{\!a}
  =3(\Psi-\kappa)-3\int_0^\tau\sigma(\tau'){\cal H}(\tau')\,
    d\tau'+A\ ,\qquad
\end{eqnarray}
where $A$ is an integration constant. Equation (\ref{denpert}) also
follows from energy conservation in a perturbed Robertson-Walker
spacetime, which gives the perturbed continuity equation
\begin{equation}\label{pertecons}
  \dot\delta+3{\cal H}\sigma=3\dot\Psi+\Delta u
\end{equation}
where $u$ is the velocity potential.  For long wavelength
perturbations of wavenumber $k\to0$, $\Delta u=-k^2u$ is generally
negligible compared with the other terms in the equation.

Determining the velocity perturbations requires considering spatial
variations of the velocity potential, which do not exist in a
perfectly homogenous Robertson-Walker spacetime.  Thus we consider a
perturbed Robertson-Walker spacetime with metric (\ref{pertrw}) and
energy-momentum tensor (\ref{emtensor}).  Evaluating $\nabla_\mu
T^\mu_{\ \,i}=0$ gives the first-order result
\begin{equation}\label{pertpcons}
  \dot u+(1-3c_w^2){\cal H}u=c_w^2\delta+\sigma+\Phi+(\Delta+3K)\pi\ .
\end{equation}
If the shear stress can be neglected, we can integrate to obtain
\begin{equation}\label{upert}
  u(\tau)=\frac{1}{(\rho+p)a^4}\int_0^\tau\left[c_w^2(\tau')
    \delta(\tau')+\sigma(\tau')+\Phi(\tau')\right]
    [\rho(\tau')+p(\tau')]a^4(\tau')\,d\tau'+\frac{B}{(\rho+p)a^4}\ ,
\end{equation}
where $B$ is an integration constant.

Equations (\ref{denpert})--(\ref{upert}) do not assume the Einstein
field equations -- they hold for any theory of gravity provided
consistent with local energy-momentum conservation.  However, the
integral forms of these equations are rarely used.  Instead, given
the potential $\Psi$ on long wavelengths, the usual procedure in
general relativity is to evaluate the density and velocity potential
using initial-value constraints, of which the Poisson equation is
one. That these constraint equations are more general will be shown
next.

We define the following combinations of metric and energy-momentum
perturbations:
\begin{mathletters}\label{constraints}
\begin{eqnarray}
  C_1&=&(\Delta+3K)\Psi-\gamma(\delta+3{\cal H}u)\ ,\ \label{C1}\\
  C_2&=&\dot\Psi+{\cal H}\Phi-\gamma u\ ,\label{C2}\\
  C_3&=&\Psi-\Phi-3\gamma\pi\ .\label{C3}
\end{eqnarray}
\end{mathletters}
Here, $\delta+3{\cal H}u$ is the number density perturbation on a
hypersurface in which $T^0_{\ \,i}$ vanishes, i.e. number the
density perturbation in the local fluid rest frame.

In general relativity each of the constraints vanishes
($C_1=C_2=C_3=0$), however we will not assume this to be
automatically true. Differentiating the constraints and using
equations (\ref{Hdot}), (\ref{pertecons}) and (\ref{pertpcons})
gives
\begin{mathletters}\label{dconstr}
\begin{eqnarray}
  &&\frac{1}{a}\frac{\partial}{\partial\tau}(aC_1)-(\Delta+3K
    -3\gamma)C_2-{\cal H}(\Delta+3K)C_3=3\gamma({\cal H}^2+K
    -\dot{\cal H}-\gamma)u\ ,\label{cons1}\\
  &&\frac{\gamma}{a}\frac{\partial}{\partial\tau}\left(
    \frac{a}{\gamma}C_2\right)-c_w^2C_1+\frac{\gamma}{{\cal H}}
    \frac{\partial}{\partial\tau}\left(\frac{{\cal H}^2}{\gamma}
    C_3\right)-KC_3\nonumber\\
  &&\qquad=\frac{\gamma}{{\cal H}}\frac{\partial}{\partial\tau}
    \left[\frac{{\cal H}^2}{\gamma a^2}\frac{\partial}{\partial\tau}
    \left(\frac{a^2}{{\cal H}}\Psi\right)\right]-c_w^2\Delta
    \Psi-\gamma\left[\sigma+\frac{3}{{\cal H}}\frac{\partial}
    {\partial\tau}\left({\cal H}^2\pi\right)+\Delta\pi\right]
    \nonumber\\
  &&\qquad+({\cal H}^2+K-\dot{\cal H}-\gamma)(\Phi-2\Psi)
    -[2\gamma-3(\eta^2+K)(1+w)]\Psi\ .
    \label{cons2}
\end{eqnarray}
\end{mathletters}
Here we set $\gamma\equiv4\pi Ga^2(\bar\rho+\bar p)$ and have
assumed nothing about gravity.  If we assume that the Einstein field
equations hold, the right-hand sides of equations (\ref{dconstr})
vanish.  If we make the weaker assumption that the background is
governed by the Friedmann equation and that $\Delta\Psi$ and $\pi$
(shear stress) can be neglected on large scales so that equation
(\ref{psidotfrw}) holds, the right-hand sides still vanish. In
keeping with the previous treatment we also assume $\Psi=\Phi$ so
that $C_3=0$. With these assumptions we can integrate equations
(\ref{dconstr}) to obtain
\begin{equation}\label{constrsol}
  C_1=3{\cal H}C_2-A\gamma\ ,\ \
  C_2=-\frac{A}{a^2}\int_0^\tau c_w^2(\tau')\gamma(\tau')a^2(\tau')\,
    d\tau'-\frac{4\pi GB}{a^2}\ ,
\end{equation}
where $A$ and $B$ are integration constants.  Comparison with
equations (\ref{denpert}), (\ref{upert}), and (\ref{constraints})
shows that these are exactly the same initial-value constants
obtained from integrating the equations of energy-momentum
conservation.  In other words, they correspond to changing $\delta$
and $u$ without changing $\Psi$. The third initial-value constant
$C$ appearing in equation (\ref{curv}) contributes nothing new to
the constraints because the pure decaying mode has
$\delta=3\Psi=-3{\cal H}u$; the decaying mode can be eliminated by a
change of the time coordinate.

In general relativity, the Einstein field equations give
$C_1=C_2=C_3=0$ implying $A=B=0$. However, in other theories of
gravity the constraints may be nonzero without contradicting
equations (\ref{dconstr}). The initial-value constants follow from
\begin{equation}\label{AB}
  A=\lim_{a\to0}\frac{T\delta S}{a^3(\bar\rho+\bar p)}\ ,\ \
  B=\lim_{a\to0}a^4(\bar\rho+\bar p)u\ .
\end{equation}
While nonzero values $A$ and $B$ are testable, in principle, by
comparing the density and velocity fields of galaxies with the
gravitational potential implied by gravitational lensing or
hydrostatic equilibrium in clusters of galaxies, much more stringent
tests obtain at high redshift, where the $A$ and $B$ terms
contribute relatively more to the energy-momentum tensor.  It would
be interesting to place limits on these constants using measurements
of the cosmic microwave background anisotropy.

Assuming that $A$ and $B$ are small compared with $\delta$ and
$a^4(\bar\rho+\bar p)u$, respectively, at low redshift, and assuming
furthermore that the Friedmann equation and energy-momentum
conservation are valid, then the density and velocity fields in a
perturbed Robertson-Walker spacetime whose metric takes the form of
equation (\ref{rw3}) must on large scales obey the same constraint
equations as in general relativity, namely $C_1=C_2=C_3=0$.

Although the linearized Einstein field equations give three
constraints, the first two ($C_1=C_2=0$) are initial-value
constraints with no dynamical content. If $C_1=C_2=0$ on an initial
timeslice, energy-momentum conservation combined with the other
Einstein equations is enough to ensure $C_1=C_2=0$ for all times.
The third constraint, $C_3=0$, is a true dynamical constraint
because its time derivative is not forced to vanish as a result of
the Einstein field equations and energy-momentum conservation.

We will see later that initial-value constraints exist not only in
GR but in any gravity theory yielding a long-wavelength perturbed
Robertson-Walker solution with local energy-momentum conservation.
The key distinguishing features of general relativity are then seen
to be the Friedmann equation plus the dynamical constraint $C_3=0$.

\subsection{Multicomponent Fluids}
\label{sec:multi}

Equations (\ref{isocurv}), (\ref{denpert}) and (\ref{upert}) are
true quadratures only if $\sigma(\tau)$ is known. While equation
(\ref{sigdef}) gives $\sigma$ from $p(\rho,S)$ or $\rho(a,S)$, the
entropy depends on the internal degrees of freedom of a fluid.

Consider a multicomponent imperfect fluid whose density and pressure
are described by a set of parameters $\{X\}$ which can vary with
position at fixed expansion factor $a$.  We replace the single
entropy $S$ by as many parameters as are necessary to characterize
spatial variations in the equation of state of the fluid. For
example, a system of non-interacting fluids has
\begin{equation}\label{multifluid}
  \rho(a,X)=\sum_i\tilde\rho_ir_i(a)\ ,\ \
  p(a,X)=\sum_iw_i\tilde\rho_ir_i(a)\ ,\ \
  r_i(a)=\exp\left[3\int_a^1(1+w_i)\,
  d\ln a\right]\ ,
\end{equation}
where the $\tilde\rho_i$ are independent of $a$. The $w_i$ generally
are not spatially varying, for example $w=0$ for cold dark matter
(``dust'') and $w=\frac{1}{3}$ for a relativistic ideal gas
(``radiation''). In this case the parameters are the abundances of
each fluid component, $\{X\}=\{\tilde\rho_1,\tilde\rho_2,\ldots\}$.
The same procedure will work regardless of what parameters
characterize the multicomponent fluid but the set $\{X\}$ includes
only those that can vary with position at a fixed value of $a$ ($a$
playing the role of volume).

The net equation of state and sound speed parameters are $w\equiv
p(a,X)/\rho(a,X)$ and $c_w^2\equiv(\partial p/\partial a)/(\partial
\rho/\partial a)$.  The entropy perturbation follows similarly to
equation (\ref{sigdef}):
\begin{equation}\label{sigmulti}
  \sigma=\sum_X\frac{\delta X}{\rho+p}\left(\frac{\partial p}
    {\partial X}-c_w^2\frac{\partial\rho}{\partial X}\right)
  =-\frac{1}{3}\frac{\partial\epsilon}{\partial\ln a}\ ,\ \
  \epsilon(a,X)\equiv\sum_X\frac{\delta X}{\rho+p}\left(\frac{\partial\rho}
    {\partial X}\right)_{\!a}\ .
\end{equation}
The symbol $\epsilon$, equalling the number density perturbation at
fixed expansion factor $a$, is introduced here to avoid confusion
with the number density perturbation $\delta$ measured at fixed
$\tau$. For the example of equation (\ref{multifluid}),
\begin{equation}\label{multisig0}
  \sigma=\sum_if_i(w_i-c_w^2)\epsilon_i=\sum_if_iw_i
    (\epsilon_i-\epsilon)\ ,
\end{equation}
where
\begin{equation}\label{multidefs}
  f_i(a)\equiv\frac{(\tilde\rho_i+\tilde p_i)r_i(a)}{\rho+p}\ ,\ \
  \epsilon_i\equiv\frac{\delta\tilde\rho_i}{\tilde\rho_i+\tilde p_i}\ ,\ \
  \epsilon=\sum_if_i\epsilon_i\ .
\end{equation}
Here $\tilde p_i=w_i\tilde \rho_i$ and $\epsilon_i$ are both
independent of time and $f_i$ is the enthalpy fraction for component
$i$, normalized so that $\sum_if_i=1$. We see that the entropy
perturbation of a superposition of ideal gases arises entirely from
differences in the particle number density of the different species
measured at fixed expansion $a$. For example, for cold dark matter
(subscript $m$) and radiation (subscript $r$) with
$y(a)\equiv\rho_m/\rho_r$, $f_m=3y/(3y+4)$, $f_r=4/(3y+4)$, and
$\sigma=\frac{1}{3}f_rf_m(\epsilon_r-\epsilon_m)$. For a
superposition of $N$ fluids there are $N-1$ independent entropy
modes corresponding to the $N-1$ independent differences of the
number density perturbations $\epsilon_i$.

In the general case where the individual components have a
time-varying $w_i$ and may also have entropy perturbations,
\begin{equation}\label{multisig}
  \sigma=\sum_if_i\left[\sigma_i+(c_i^2-c_w^2)\epsilon_i\right]
    =\sum_if_i\left[\sigma_i+c_i^2(\epsilon_i-\epsilon)\right]\ ,
\end{equation}
where $c_i^2\equiv[\partial(w_ir_i)/\partial a]/(\partial r_i/
\partial a)$ and $\sigma_i$ are the sound speed squared and entropy
perturbation for each component, respectively.  As an example,
consider a tightly-coupled plasma of photons and nonrelativistic
baryons with $w=[3(1+y_b)]^{-1}$ where $y_b(a)\equiv\rho_b/\rho_r$
(subscript $b$ denotes baryons with $r$ for the photons).  In this
case, $w=f_r/(4-f_r)$ and $c_w^2=\frac{1}{3}f_r$ compared with
$w=f_r/(4-f_r)$ where $f_r=4/(3y_b+4)$. Although this plasma behaves
like a single fluid, it can have a nonzero entropy perturbation
arising from initial variations in the baryon to photon ratio,
$\sigma=\frac{1}{3}f_rf_b(\epsilon_r-\epsilon_b)$ where
$\epsilon_r-\epsilon_b=\delta\ln(n_r/n_b)$, the fractional
perturbation in the photon to baryon number density ratio, is
constant in time. Thus, by examining the composition of each fluid
component, we can write the time-dependent total entropy
perturbation $\sigma$ in terms of a set of time-independent
constants $\epsilon_i$.  By doing so, we are able to fully reduce to
quadratures the metric perturbation as shown in \S \ref{sec:quadGR}.

Using the same methods we obtain exact results for the density of
individual fluid components.  Here one must be careful to evaluate
the perturbations at fixed $(\tau,\chi)$ after the coordinate
transformations $\tau\to\tau+\alpha(\tau)$ and
$\chi\to\chi(1-\kappa)$ discussed in \S \ref{sec:cpert}.  The
density perturbation of component $i$ at fixed $\tau$ is
\begin{equation}\label{dpert}
  \delta_i(\tau)=\frac{\rho_i(a(\tau+\alpha,K+\delta K,X+\delta X),
    X+\delta X)-\rho_i(a,X)}{(\rho+p)_i}=3(\Psi-\kappa)+\epsilon_i\ .
\end{equation}
The $\epsilon_i$ comes from the initial number density perturbation
at fixed $a$ while $3(\kappa-\Psi)$ is the fractional change in
volume introduced by shifting the initial hypersurface of constant
$\tau$.  Averaging over all components with weights $f_i$ gives the
net density perturbation
\begin{equation}\label{dpertnet}
  \delta(\tau)=3(\Psi-\kappa)+\epsilon(\tau)\ ,
\end{equation}
where $\epsilon(\tau)\equiv\epsilon(a(\tau,K,X),X)$ is given by
equation (\ref{sigmulti}) and we are holding $K$ and $X$ fixed to
lowest order in perturbation theory. By comparing equations
(\ref{denpert}) and (\ref{dpertnet}) and using
$\partial\epsilon/\partial\ln a=-3\sigma$ we see that the constant
$A$ has been determined.

The velocity potential of individual components follows from
equation (\ref{upert}) with subscript $i$ applying to all quantities
except $a$, $\tau$, and $\Phi$.  For $N$ uncoupled fluids there are
$N$ constants $B_i$.  The net velocity potential is the average over
the fluids, $u=\sum_if_iu_u$.

We have succeeded in reducing the long-wavelength density, velocity,
and entropy perturbations of all fluid components to quadratures.
This presentation works for both general relativity and alternative
theories of gravity.  In the former case, equations (\ref{curv}) and
(\ref{isocurv}) reduce the problem entirely to quadratures specified
by a set of constants
$(\kappa,\epsilon_1,\ldots\epsilon_N,B_1,\ldots,B_N)$.  We explore
next the reduction of the metric perturbation to quadratures for
alternative theories of gravity.

\section{Quadratures for Alternative Theories of Gravity}
\label{sec:alt}

Much of the treatment given so far assumes the validity of the
Friedmann equation or GR.  It is straightforward to redo the
calculations of equations (\ref{rw2})--(\ref{psikappa}) and
(\ref{aderivS})--(\ref{psidotfrw}) without making these assumptions.
The expansion rate ${\cal H}$ can depend only on the scale factor
$a$ and curvature $K$ of the background Robertson-Walker spacetime
and the parameters $\{X\}$ describing the equation of state as
discussed in \S \ref{sec:multi}: ${\cal H}= {\cal H}(a,K,X)$.  In
this case equation (\ref{friedsol}) is replaced by some function
$\tau(a,K,X)=\int_0^a[a{\cal H}(a,K,X)]^{-1}\,da$.  In addition, for
an arbitrary theory of gravity the scalar mode has two distinct
gravitational potentials, equation (\ref{pertrw}).  Repeating the
derivations of \S\S \ref{sec:cpert} and \ref{sec:epert} with the
metric of equation (\ref{pertrw}) with $\Phi\ne\Psi$ and with an
arbitrary background expansion rate ${\cal H}(a,K,X)$ gives the
following result for long-wavelength metric perturbations with
$\pi=0$:
\begin{eqnarray}\label{pside}
  \frac{1}{a^2}\frac{\partial}{\partial\tau}\left(\frac{a^2\Psi}
    {{\cal H}}\right)+(\Phi-\Psi)&=&\frac{\kappa}{a}\frac{\partial}{\partial
    \tau}\left(\frac{a}{{\cal H}}\right)+{\cal H}\frac{\partial}
    {\partial\ln a}\left[2\kappa\left(\frac{\partial\tau}
    {\partial\ln K}\right)_{\!a,X}+\sum_X\delta X\left(\frac{\partial\tau}
    {\partial X}\right)_{\!a,K}\,\right]\nonumber\\
  &=&\frac{\kappa}{a}\frac{\partial}{\partial
    \tau}\left(\frac{a}{{\cal H}}\right)-2\kappa\left(\frac{\partial
    \ln{\cal H}}{\partial\ln K}\right)_{\!a,X}-\sum_X\delta X\left(
    \frac{\partial\ln{\cal H}}{\partial X}\right)_{\!a,K}\ .
\end{eqnarray}
The quantities $\kappa$ and $\delta X$ are (for long wavelengths)
independent of time.

Equation (\ref{pside}) provides one differential equation for two
functions $\Phi$ and $\Psi$.  Without a second relation, provided by
a law of gravity, we can determine neither $\Phi$ nor $\Psi$.

Even without a full theory of gravity, however, we can still obtain
interesting and useful results if we assume energy-momentum
conservation.  In this case we have the quadratures (\ref{upert})
and (\ref{dpertnet}).  By combining these equations with equation
(\ref{pside}) one can derive the following results which generalize
the energy-momentum constraints $C_1$ and $C_2$ of general
relativity in the limit of vanishing shear stress ($\pi=0$) and long
wavelengths (the spatial Laplacian $\Delta=0$ when applied to all
perturbations):
\begin{mathletters}\label{genconstraints}
\begin{eqnarray}
  \delta+3{\cal H}u&=&\frac{3{\cal H}B'}{a^4(\rho+p)}\ ,\ \label{C1gen}\\
  \dot\Psi+{\cal H}\Phi&=&-a\frac{\partial}{\partial\tau}\left(
    \frac{{\cal H}}{a}\right)u+\frac{{\cal H}}{a^2(\rho+p)}\frac{\partial}
    {\partial a}\left(\frac{{\cal H}B'}{a}\right)\ ,\label{C2gen}
\end{eqnarray}
\end{mathletters}
where
\begin{equation}\label{Bdef}
  B'(a)\equiv\int_0^a a^2(\rho+p)
    \left[-2\kappa\left(\frac{\partial\ln{\cal H}}{\partial\ln K}
    \right)_{\!a,X}-\sum_X\delta X\left(\frac{\partial\ln{\cal H}}
    {\partial X}\right)_{\!a,K}+\frac{\epsilon}{3a}
    \frac{\partial}{\partial\tau}\left(\frac{a}{{\cal H}}\right)
    _{\!K,X}\,\right]\,\frac{ada}{{\cal H}}+B\ .
\end{equation}
The constant $B$ is essentially the same integration constant as in
equation (\ref{upert}).  Without loss of generality we may set $B=0$
in equation (\ref{Bdef}).

Equations (\ref{dpertnet})--(\ref{genconstraints}) are not all
independent.  Any three of the four imply the remaining one.  Thus,
we may describe the perturbations by equations (\ref{dpertnet}) and
(\ref{genconstraints}).  To reduce the system fully to quadratures
requires one more equation equivalent to a relation between $\Phi$
and $\Psi$.

It is worth emphasizing that the reduction to quadratures is based
on only a few assumptions:
\begin{enumerate}
  \item The spacetime is nearly Robertson-Walker with perturbation
    amplitude small enough for linear perturbation theory to apply.
  \item Local energy-momentum conservation holds: $\nabla_\mu
    T^{\mu\nu}=0$.
  \item In the conformal Newtonian gauge, spatial gradients of
    fields are small enough to be neglected in all equations of
    motion.
  \item Shear stress perturbations are negligible.
  \item A theory of gravity provides some relation between $\Phi$
    and $\Psi$ and possibly the matter perturbations.  This
    assumption is needed only for a complete reduction to quadratures.
\end{enumerate}
There are no restrictions on the equation of state of matter and
radiation fields.  There are no restrictions on the geometry of
perturbations; in particular, no assumption of spherical symmetry.
The neglect of spatial gradients becomes invalid on scales over
which non-gravitational forces act so as to modify local
energy-momentum conservation.  By definition, spatial gradients of
pressure are important and modify our results on scales less than
the Jeans length.

The constraints (\ref{genconstraints}) are particularly useful when
$B'=0$.  For any theory having a flat Robertson-Walker solution
(including GR), $\lim_{K\to0}(\partial{\ln{\cal H}}/\partial\ln
K)=0$.  When there are no entropy perturbations, $\epsilon=0$. Thus,
in a flat universe with initially isentropic perturbations, the only
possible contribution to $B'$ comes from the $\sum_X\delta
X(\partial\ln{\cal H}/\partial\ln X)$ term.  If the only dependence
of ${\cal H}$ on equation of state parameters $X$ is through the
density $\rho$, then
\begin{equation}\label{Hrho}
  \sum_X\delta X\left(\frac{\partial\ln{\cal H}}{\partial X}\right)_{\!a,K}
  =\epsilon(\rho+p)\frac{\partial\ln{\cal H}}{\partial\rho}\ ,
\end{equation}
which vanishes for isentropic initial conditions $\epsilon=0$. Thus,
for a broad class of theories, $B'=0$ for curvature perturbations in
a flat universe implying the long-wavelength constraints
$\delta+3{\cal H}u=0$, $\dot\Psi+{\cal H}\Phi=\gamma u$ where
$\gamma=-a\partial({\cal H}/a)/\partial\tau$ for $K=0$. From
equation (\ref{dpertnet}) we also have $\delta=3(\Psi-\kappa)$ for
long wavelengths and isentropic initial conditions.  In this case we
have three relations between the four time-dependent functions
$(\Phi,\Psi,\delta,u)$.  In the general case we have an additional
function, $B'$, whose specification requires a theory for the
homogeneous expansion.

Thus, without knowing anything about the underlying gravity theory
except that it is consistent with local energy-momentum conservation
and that it admits a Robertson-Walker solution, we have deduced
initial-value constraints similar to those of GR under the
conditions of long wavelengths and negligible shear stress.  One is
not free to specify arbitrary metric and matter perturbations on
long wavelengths, at least under the assumption that long-wavelength
perturbations evolve like separate universes.

However, the initial-value constraints (\ref{genconstraints})
combined with equations (\ref{sigmulti}) and (\ref{dpertnet}) are
insufficient to fully determine the growth of perturbations.  We
need a third constraint, a relation between $\Phi$ and $\Psi$ given
by some theory of gravity. Such a dynamical constraint would allow
equation (\ref{pside}) to be integrated and, with the initial-value
constraints, reduce $(\Phi,\Psi,\delta,u)$ fully to quadratures for
long-wavelength perturbations with vanishing shear stress.

The simplest case of a dynamical constraint on the gravitational
fields is $\Phi=\Psi$ for $\pi=0$ as in GR (although this is more
general than GR since we do not assume the Friedmann equation).
Under this assumption, equation (\ref{pside}) may be integrated
completely giving (suppressing the dependencies on $K$ and $X$ and
using $a$ as the time variable)
\begin{equation}\label{general}
  \Psi(a)=\kappa\Psi_+(a)+\sum_X(\delta X)\Psi_X(a)+C\Psi_-(a)
  \ \ \hbox{if}\ \ \Phi=\Psi\ ,
\end{equation}
where
\begin{equation}\label{curvgen}
  \Psi_+(a)=\frac{{\cal H}}{a^2}\int_0^a\left[1-\left(\frac{\partial
    \ln{\cal H}}{\partial\ln a}\right)_{\!K,X}-2\left(\frac{\partial
    \ln{\cal H}}{\partial\ln K}\right)_{\!a,X}\,\right]\,\frac{ada}{{\cal H}}
    \ ,\ \
  \Psi_-(a)={\cal H}a^{-2}\ ,
\end{equation}
and
\begin{equation}\label{entgen}
  \Psi_X(a)=-\frac{{\cal H}}{a^2}\int_0^a\left(\frac{\partial\ln{\cal H}}
    {\partial X}\right)_{\!a,K}\,\frac{ada}{{\cal H}}.
\end{equation}
The potential has been reduced entirely to quadratures depending
only on the expansion rate ${\cal H}$ as a function of expansion,
curvature, and equation of state parameters.

Considering again a flat universe with isentropic initial conditions
and assuming equation (\ref{Hrho}) holds, but now with the added
condition $\Phi=\Psi$, we see that $\Psi=\kappa\Psi_+$ is determined
completely by the expansion rate ${\cal H}(a,0,X)$ with constant
values of the equation of state parameters $X$.  The density and
velocity potential perturbation follow from $\delta=3(\Psi-\kappa)=
-3{\cal H}u$. In this case the growth of long-wavelength
perturbations is completely determined by the expansion history
${\cal H}(a)$ with fixed curvature ($K=0$) and composition ($\delta
X=0$).

For example, in the DGP brane-world theory of Dvali et al.\ (2000),
the Friedmann equation is modified by replacing $\rho$ by
$(\sqrt{\rho+\rho_{r_c}}+\sqrt{\rho_{r_c}})^2$ with $\rho_{r_c}$
being a constant. If $\Phi=\Psi$, equation (\ref{psikappa}) remains
valid for the long-wavelength curvature mode provided that $\gamma$
changes to
\begin{equation}\label{dgpgamma}
  \gamma'\equiv{\cal H}^2+K-\dot{\cal H}=\frac{3}{2}(1+w)
    ({\cal H}^2+K)\left(\frac{\sqrt{\rho+\rho_{r_c}}}
    {\sqrt{\rho+\rho_{r_c}}+\sqrt{\rho_{r_c}}}\right)\ .
\end{equation}
The isocurvature mode is also modified slightly, with
$\gamma\to\gamma'\rho/(\rho+\rho_{r_c})$ in the last line of
equation (\ref{psidotfrw}). Equations (\ref{denpert}) and
(\ref{upert}) are unchanged.

The solutions implied here are only valid as $k\to0$ so that spatial
gradient terms may be neglected in the equations of motion.  The DGP
model has a length scale $r_c=(32\pi G\rho_{r_c}/3)^{-1/2}$
comparable to $H_0^{-1}$.  The long-wavelength limit becomes
$kr_c\ll1$ suggesting that the formal solution given here is not
applicable for wavelengths shorter than the present Hubble length.
Other theories of gravity will give different ranges of
applicability.  In general, any long-range departures from general
relativity will modify the evolution of patches of a
Robertson-Walker spacetime.

The results of this section highlight the importance of testing the
dynamical constraint $C_3=\Psi-\Phi-3\gamma\pi=0$ in general
relativity.  The relation between $\Phi$ and $\Psi$, combined with
the background expansion rate ${\cal H}(a,K,X)$, is the key to
long-wavelength perturbations of a Robertson-Walker cosmology.

\section{Comparing Geometric and Dynamic Methods for Testing Gravity
  and Dark Energy}

As we have seen, the evolution of long-wavelength perturbations
requires two ingredients: (1) a relation between the two
gravitational potentials $\Phi$ and $\Psi$ (possibly involving
auxiliary fields), and (2) a generalization of the Friedmann
equation for the background expansion rate.  Without the first
ingredient, the long-wavelength perturbations cannot be predicted.
Let us assume that a dynamical constraint exists relating $\Phi$ and
$\Psi$ so that equation (\ref{pside}) can be integrated.  Then the
evolution of perturbations is determined by the expansion rate of
the universe, the very same information that determines the
redshift-distance relation used by geometric methods.  Does this
mean that dynamic and geometric methods measure the same thing?

Not necessarily.  First, the dependence of perturbation (dynamic)
methods on the expansion rate necessarily depends on the value of
$\Phi-\Psi$ which may be nonzero for theories of gravity other than
GR.  Second, dynamic methods depend not only on ${\cal H}(a,K,X)$
and its variation with $a$ but also on $(\partial\ln{\cal
H}/\partial\ln K)_{a,X}$ (for curvature perturbations) or
$(\partial\ln{\cal H}/\partial X)_{a,X}$ (for entropy perturbations;
here $X$ are parameters describing the fluid composition or equation
of state). Geometric methods, by comparison, depend only on ${\cal
H}(a,K,X)$ with fixed $K$ and $X$.  If the universe is nonflat or
there exist entropy perturbations, dynamic methods have the
potential to reveal information about the dependence of ${\cal H}$
on $K$ and $X$ which is not present in the expansion history for one
universe with fixed $K$ and $X$.  Finally, the quadrature results
for perturbations are valid only in the limit of long wavelengths;
long-range forces different from Einstein gravity may lead to
spatial gradient terms that modify the quadratures on presently
observable scales.

Geometric tests are generally cast in terms of the luminosity
distance or angular-diameter distance but in fact depend strictly on
the redshift-distance relation for radial null geodesics,
\begin{equation}\label{lightcone}
  \chi(z)=\tau_0-\tau(a,K,X)=\int_0^z\frac{dz'}{H(z')}\ ,
\end{equation}
where $\tau_0=\tau(1,K,X)$ and $a=(1+z)^{-1}$. The
redshift-dependence of the long-wavelength curvature perturbation
$\Psi_+(z)$ requires specifying a theory of gravity. If $\Phi=\Psi$,
for both general relativity and DGP brane worlds the same result
holds on very long wavelengths:\footnote{This does not mean that the
two theories predict identical perturbations on scales larger than
the Jeans length.  The DGP model has a much longer length scale,
$r_c$, below which equation (\ref{zint}) may be invalid.}
\begin{equation}\label{zint}
  \Psi_+(z)=(1+z)H(z)\int_z^\infty\left[\frac{K}{H^2(z')}
    +\frac{1}{(1+z')H(z')}\frac{dH}{dz'}\right]\,
    \frac{dz'}{H(z')}\ .
\end{equation}

More generally, let us make no assumptions about $\Phi-\Psi$ but
consider a flat universe with negligible shear stress and initially
isentropic curvature fluctuations as predicted by the simplest
inflationary universe models.  In this case, equation (\ref{pside})
reduces to the following relation between $H(z)$, $\Psi(z)$, and
$\Phi(z)$ for long wavelengths:
\begin{equation}\label{HPsiPhi}
  \frac{\Phi(z)}{1+z}=\frac{d\Psi}{dz}+\frac{(\kappa-\Psi)}
    {H}\frac{dH}{dz}\ .
\end{equation}
Here $\kappa$ is independent of redshift on large scales although
$\kappa$, $\Psi$, and $\Phi$ will vary with position or spatial
wavenumber.  In principle, measurements of $H(z)$ from geometric
methods and $\Psi(z)$ from perturbations could (for isentropic
perturbations of a flat universe with negligible shear stress)
determine $\Phi(z)$ up to an additive term proportional to $d\ln
H/d\ln(1+z)$ (since the relation between $\kappa$ and $\Psi$ is
unknown if the theory of gravity is unspecified).  In particular,
such measurements could enable GR or alternative theories of gravity
to be tested without any assumptions about the density and pressure
of mass-energy in the universe.

This kind of test is difficult to imagine carrying out because of
the difficulty of measuring $\Psi(z)$.\footnote{Observations of
density perturbations, e.g.\ from galaxy redshift surveys, are
easier than observations of the gravitational potential
perturbations.  The relation between density perturbations and
$\Psi$ depends on the theory of gravity and may require specifying
more than just $H(z)$.} Alternatively, if one assumes that GR is
valid, dynamic methods can be used to provide additional constraints
on dark energy because the geometric and dynamic methods have a
different dependence on the equation of state of dark energy.

\begin{figure}[t]
  \begin{center}
    \includegraphics[scale=0.6]{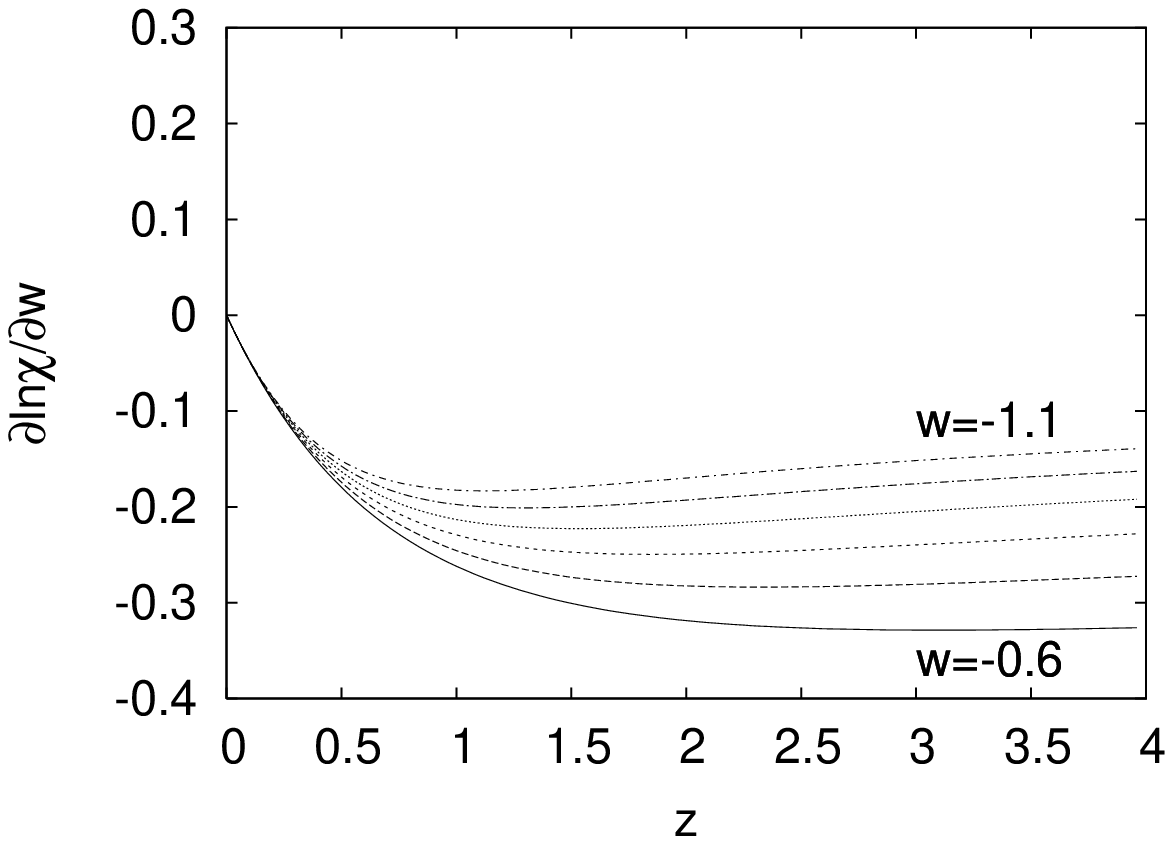}
    \includegraphics[scale=0.6]{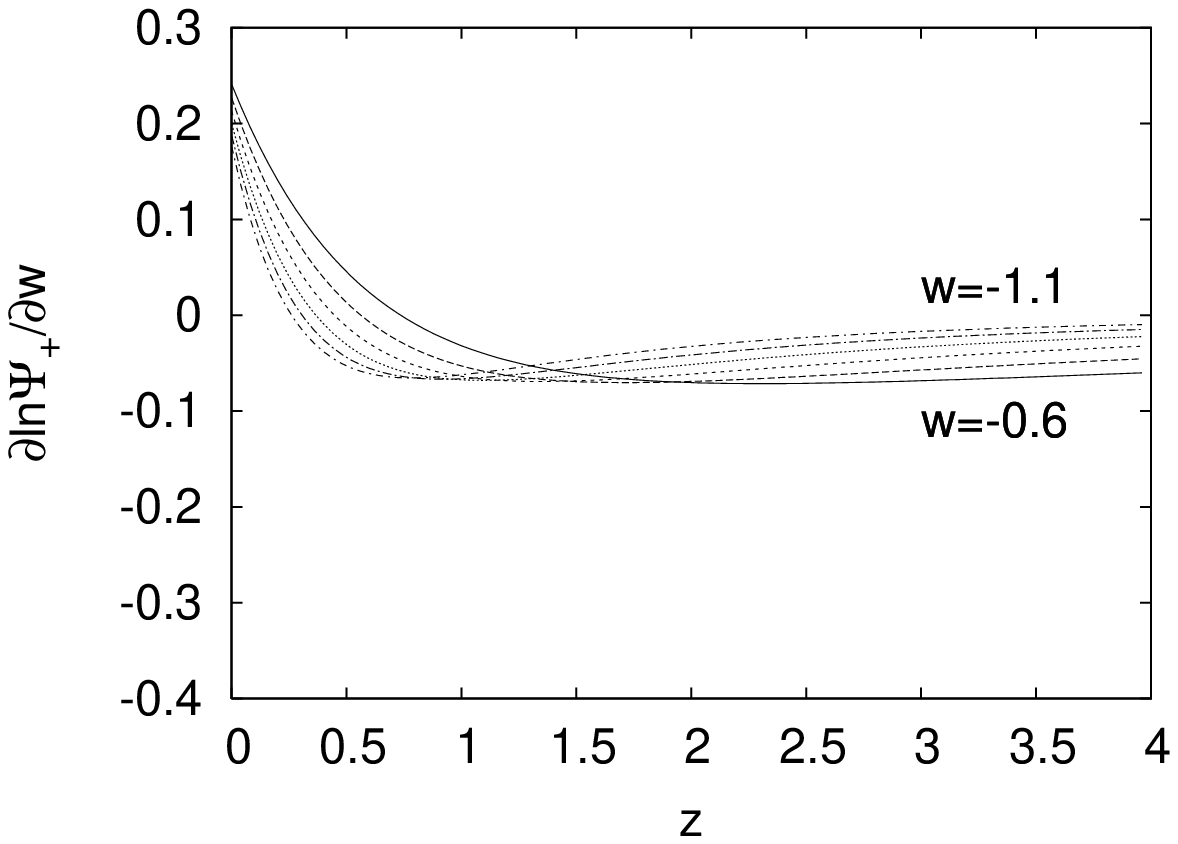}
  \end{center}
  \caption{Sensitivity of geometric methods (left panel, characterized
  by the comoving distance $\chi$ to redshift $z$) and dynamic
  methods (right panel, characterized by the curvature perturbation
  $\Psi_+$ at redshift $z$) to the equation of state parameter $w$.
  A flat model with $\Omega_m=0.3$ was assumed.  Curves are shown
  for $w=-0.6,-0.7,\ldots,-1.1$.  Dynamic (perturbation) methods are
  insensitive to dark energy at high redshift but are more sensitive
  than geometric methods at low redshift.}
    \label{fig:fig1}
\end{figure}

Geometric measurements of $\chi(z)$ at a given redshift depend on
the expansion history at smaller redshifts; dynamic measurements of
$\Psi(z)$ depend on $H(z)$ at higher redshifts.  Because the effects
of dark energy typically decline with increasing redshift,
perturbation methods are at a disadvantage at high redshift but may
be superior to geometric methods at low redshift.

To assess their relative merits for measuring dark energy (assuming
GR is the correct gravity theory) the methods were compared using a
flat model with $\Omega_m=0.3$ and a dark energy component with
constant $w$ independent of $z$. While unphysical, this model is
commonly used to compare theories with data.  The logarithmic
derivatives of equations (\ref{lightcone}) and (\ref{zint}) with
respect to $w$ at fixed $z$ were evaluated numerically to produce
the results shown in Figure \ref{fig:fig1}. A large value of
$|\partial\ln\chi/\partial w|$ indicates that the geometric method
is relatively powerful --- a given measurement error in $\ln\chi$
translates into a smaller uncertainty in $w$ for larger values of
$|\partial\ln\chi/\partial w|$.  The same is true for dynamic
methods using $\ln\Psi_+$.

As expected, perturbation methods are insensitive to the dark energy
parameter $w$ at high redshift.  At low redshift, dark energy
dominates more rapidly for smaller $w$ leading to a greater
suppression of linear growth hence a larger variation with $w$.
Geometric methods, on the other hand, are insensitive to $w$ at
small redshift where the lookback time is much less than the Hubble
time.  For the standard Friedmann equation and $w\approx-1$, dynamic
methods are more sensitive than geometric methods for $z<0.2$.  At
high redshift the comoving distance remains sensitive to $w$ despite
the declining importance of dark energy because $\chi(z)$ is an
integral over the past lightcone. The expansion history at low
redshift affects the redshift-distance relation at high redshift.

The theoretical sensitivities shown in Figure \ref{fig:fig1} must be
combined with the measurement uncertainties of the two methods
before a reliable estimate can be made of their relative merits. The
perturbation methods face a severe challenge --- to achieve a
discriminating power of $0.1$ in $w$ requires 2\% accuracy in
measurement of $\Psi_+$ at low redshift.

\section{Summary and Outlook for Cosmological Tests of Gravity}

Long-wavelength cosmological perturbations involve at least two
metric perturbations $(\Phi,\Psi)$ and two matter perturbations
$(\delta,u)$ (density and velocity potential, all in conformal
Newtonian gauge).  One might think that, without a theory of
gravity, no predictions can be made about relations between these
variables.

In fact, we have shown that, with minimal assumptions, on large
scales there are automatically three independent relations between
$(\Phi,\Psi,\delta,u)$. For an arbitrary theory of gravity these are
any three of the four equations (\ref{dpertnet}), (\ref{pside}) and
(\ref{genconstraints}a,b). Equations (\ref{genconstraints}) enforce
local energy-momentum conservation. They generalize the
initial-value constraints of general relativity. GR also has a
dynamical constraint on $\Phi-\Psi$ enabling a complete reduction to
quadratures. These relations depend on the expansion rate of the
background spacetime and on the dependence of the expansion rate on
spatial curvature (if $K\ne0$) and on the composition or entropy of
the matter filling space (if there are composition or entropy
perturbations).

Thus, the key ingredients needed to specify long-wavelength
perturbations are (1) a relation between the two gravitational
potentials $\Phi$ and $\Psi$, and (2) a relation between the
expansion rate, density, and pressure (e.g.\ the Friedmann
equation).  Given these two ingredients, we have shown how to reduce
$(\Phi,\Psi,\delta,u)$ to quadratures by introducing conserved
curvature and entropy variables.  Explicit expressions for the
time-dependence of the metric perturbations were given for Einstein
gravity, equations (\ref{curv}) and (\ref{isocurv}) or
 (\ref{general})--(\ref{entgen}).

In some circumstances (e.g. isentropic fluctuations in a flat
universe with $\Phi=\Psi$) the long-wavelength growth of
perturbations is determined completely by the expansion history
$H(z)$.  In such cases measurements of perturbation growth cannot be
combined with geometric measurements (e.g.\ supernova distances) to
discern whether dark energy is a new form of mass-energy or a
failure of the Friedmann equation.  A loophole exists in this
argument if $\Phi\ne\Psi$ or there are significant long-range
non-gravitational forces.  For a flat universe with curvature
fluctuations, the argument can be inverted to yield information
about $\Phi(z)$ from measurements of $H(z)$ and $\Psi(z)$ thereby
providing a test of gravity theories independently of assumptions
about the energy density and pressure.

The long-wavelength limit corresponds to wavelengths larger than any
relevant spatial scales so that spatial gradients may be neglected
in the equations of motion.  Pressure forces modify the dynamics on
scales smaller than the Jeans length; the dark energy may plausibly
be a scalar field with a Jeans length comparable to the Hubble
length. By measuring the wavelength-dependence of the linear growth
rate on scales greater than 1 Gpc one might measure the
time-dependence of the dark energy Jeans length and thereby
constrain its intrinsic properties. This measurement is exceedingly
difficult because the amplitude of density perturbations in the dark
energy is expected to be less than about $10^{-4}$.

In addition to providing tests of general relativity, perturbation
measurements can provide constraints on dark energy under the
assumption that general relativity is valid.  By comparing the
dependence of curvature perturbations and the redshift-distance
relation on the dark energy equation of state parameter we verified
quantitatively the expected conclusion that perturbation methods are
most useful at small redshift when the accelerated expansion begins
to suppress curvature perturbations.  A changing gravitational
potential generates cosmic microwave background anisotropy through
the integrated Sachs-Wolfe effect.  Measurement of this effect
(Padmanabhan et al.\ 2005) has the potential to further constrain
dark energy (Pogosian et al.\ 2005).

The analysis in this paper shows the importance of testing the
equality of the two Newtonian gauge gravitational potentials,
$\Phi=\Psi$ in equation (\ref{pertrw}). While this equality holds in
general relativity (in the absence of large shear stress), it may
not be true for other theories of gravity.  In addition to tests
combining geometric and perturbation methods using equation
(\ref{HPsiPhi}), one can in principle measure $\Phi-\Psi$ by
comparing the deflection of light by gravitational lenses (an effect
proportional to $\Psi+\Phi$) with the non-relativistic motion of
galaxies (an effect proportional to $\Phi$).  A similar test exists
on solar-system scales (or in binary pulsars) where the deflection
(or Shapiro delay) of light is compared with Newtonian dynamics.
Thus, for testing modified gravity as an alternative to GR it is
important to extend tests of the post-Newtonian parameter
$\gamma_{PPN}\equiv\Psi/\Phi$ (Will 2006).  Stringent limits on
$|\gamma_{PPN}-1|$ apply on solar system scales.  It would be
worthwhile to improve the limits on Mpc and larger scales by
combining weak gravitational lensing and galaxy peculiar velocity
measurements or by adding $\gamma_{PPN}$ to the parameters used in
analyzing cosmic microwave background anisotropy.

I thank Eric Linder for helpful comments.  This work was supported
by the Kavli Foundation and by NSF grant AST-0407050.

\end{document}